\documentclass[letterpaper]{jpconf}
\usepackage{graphicx}

\begin{document}
\title{New Results From CLEO and BES}

\author{Kamal K. Seth}

\address{Northwestern University, Evanston, IL, 60208, USA}

\ead{kseth@northwestern.edu}

\begin{abstract}
Latest experimental results from BES in the charmonium mass region, and those from CLEO in the bottomonium and charmonium spectroscopy are reviewed.
\end{abstract}

\section{Preamble}

I am going to talk about new results from CLEO and BES, two detectors half a world apart, trying to catch the same particles.

CLEO used to do physics in the b-quark region ($\sqrt{s} = 9-11$ GeV).  It has many credits in bottomonium ($b\bar{b}$) spectroscopy, and $B(b\bar{n}, \bar{b}n$) physics.  The CESR accelerator and the CLEO detector have now been modified to do physics in the c-quark region ($\sqrt{s} = 3-5$ GeV); this means CLEO-c for charmonium ($c\bar{c}$) spectroscopy, and $D(c\bar{n}, \bar{c}n$) physics. I will mainly talk about results in charmonium spectroscopy from CLEO and CLEO-c, but I will also present some spectroscopic results for bottomonium.

BES has worked primarily in charmonium spectroscopy since the beginning of the BEPC project.  BES was recently improved to BES II, and has logged a formidable number of $J/\psi$ and  $\psi'$.  The new results come from these datasets.

To put the physics of the $q\bar{q}$ quarkonium systems in perspective, I illustrate the charmonium and bottomonium spectra in Fig. 1.  I want to note that in the case of charmonium before now the big gaps were in the identification of the singlet states $\eta_c'(2^1S_0)$ and $h_c(1^1P_1)$, and in states above the $D\bar{D}$ threshold.  This is where remarkable progress has been made, and I will talk about it.  In the case of bottomonium there has been progress in precision, and also in identification of a new state and a new transition.  BES has reported some provocative results about baryon-antibaryon systems and light quark mesons as well.

\begin{figure}
\includegraphics[width=5.5in]{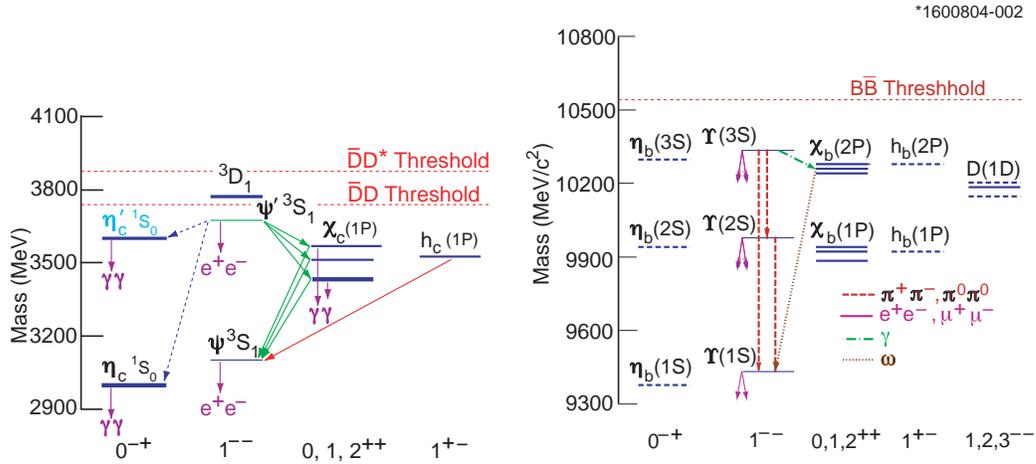}
\caption{Spectra of quarkonium states, (left) charmonium, (right) bottomonium.}
\end{figure}

The existing databases are illustrated in Fig. 2.  In the charmonium region BES dominates with 58 million $J/\psi(3097)$ and $\sim14$ million $\psi'(3686)$, but already CLEO-c has more $\psi''(3770)$.  In the bottomonium region, CLEO has greatly improved on its own earlier data, with, for example $\sim28$ million $\Upsilon(1S)$.

\begin{figure}[b]
\begin{center}
\rotatebox{-90}{\scalebox{0.4}{\includegraphics*{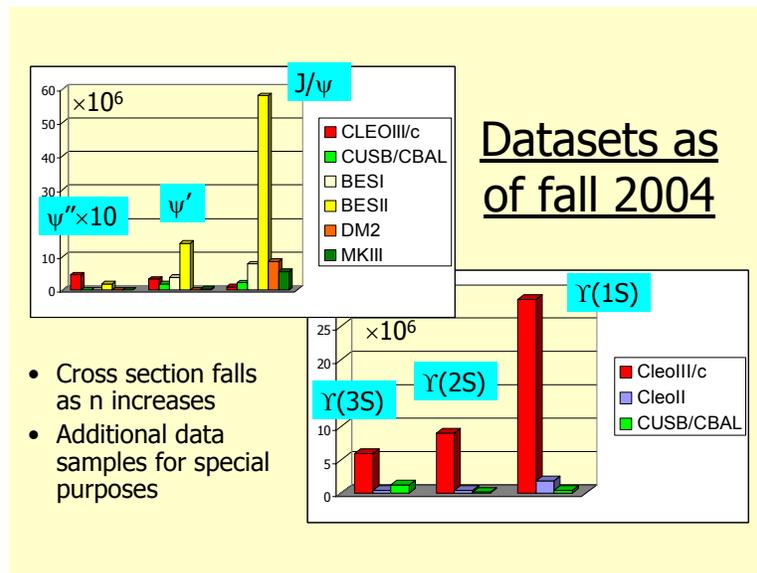}}}
\end{center}
\caption{Present datasets for $J/\psi$, $\psi'$, and $\Upsilon(1S,2S,3S)$.}
\end{figure}

\section{New Results in Charmonium Spectroscopy}

\subsection{Observation of $\eta_c'(2^1S_0)$}

The search of $\eta_c'(2^1S_0)$, the hyperfine partner of $\psi'(2^3S_1)$, has a long history.  Its identification was claimed at a mass $M(\eta_c')=3594\pm5$ MeV by the Crystal Ball in 1982 \cite{cball}, but it was not confirmed by any other experiments, including Fermilab $p\bar{p}$ experiments E760/E835 \cite{e760e835}.  Recently, Belle [3,4] reported $\eta_c'$ observation in two different measurements, with the results:
\begin{center}
\begin{tabular}{lll}
$B\to K^+\eta_c'\;,\;\eta_c'\to K_S K\pi$, & $M(\eta_c')=3654\pm6\pm8$ MeV & \cite{bellea} \\
$e^+e^-\to J/\psi \eta_c'$, & $M(\eta_c')=3622\pm12$ MeV & \cite{bellea'} \\
\end{tabular}
\end{center}
Despite the large difference between the two masses, it was clear that $M(\eta_c')$ is considerably larger than that claimed by the Crystal Ball.  To confirm the new identification, and to determine the mass more precisely, CLEO analyzed its samples of data from CLEO II and CLEO III at $\Upsilon(4S)$ to identify $\eta_c'$ in the two-photon fusion reaction $\gamma\gamma\to\eta_c'\to K_SK\pi$.  The results \cite{cleoa} are shown in Fig. 3.  The $\eta_c'$ mass was determined to be $M(\eta_c')=3642.9\pm3.1\pm1.5$ MeV.  In a similar measurement BaBar \cite{babara} obtained $M(\eta_c')=3630.8\pm3.4\pm1.0$ MeV.  The average of all these measurements is $M(\eta_c')=3637.4\pm4.4$ MeV, so that the hyperfine splitting $\Delta M_{hf}(2S)=48.6\pm4.4$ MeV.  Since $\Delta M_{hf}(1S)=115.1\pm2.0$ MeV, this result is rather surprising, and at variance with most theoretical predictions.  Is this large decrease in hyperfine splitting from $1S$ to $2S$ due to channel mixing, or does it indicate a non-scalar component in the $q\bar{q}$ confinement interaction?  This is something for theorists to work on.

\begin{figure}
\centerline{\includegraphics[width=3.in]{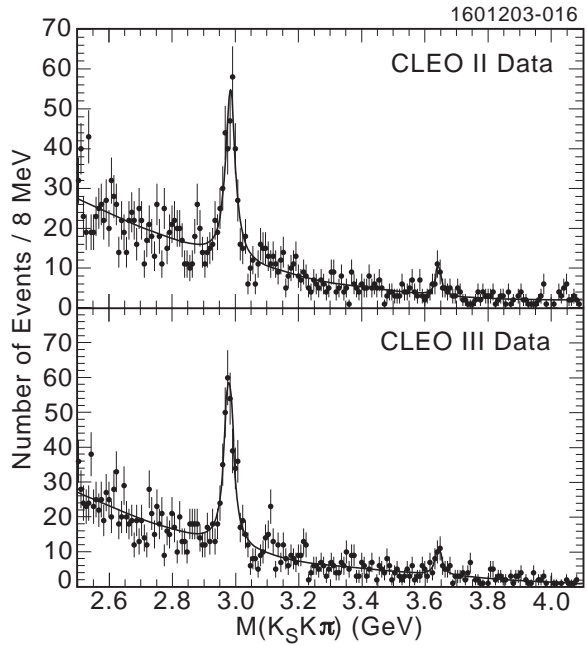}}
\caption{CLEO discovery of $\eta_c'(2^1S_0)$ in two photon fusion reaction, $\gamma\gamma\to K_S K^\pm\pi^\mp$.}
\end{figure}

\subsection{Observation of $h_c(1^1P_1)$}

The singlet $P$ resonance of charmonium $h_c(1^1P_1)$ has been much sought after, particularly because it would allow us to determine P-wave hyperfine splitting $\Delta M_{hf}(1P)\equiv\left<M(^3P_J)\right>-M(^1P_1)$, which is predicted to be zero for scalar confinement, but may have a larger value if the confinement character is different.  The Crystal Ball collaboration searched for $h_c$ in the isospin violating reaction $\psi'\to\pi^0 h_c,\;h_c\to\gamma\eta_c$ and failed to identify $h_c$ \cite{cballb}.  The Fermilab experiment E760 searched for $h_c$ in the reaction $p\bar{p}\to h_c\to\pi^0J/\psi$, and claimed to have found a signal at mass $3526.2\pm0.3$ MeV \cite{e760a}, but has failed to confirm it in a three times larger luminosity search in the same reaction in the successor experiment E835 \cite{davethesis}.  Finally, however, the successful identification of $h_c$ has just been announced by CLEO \cite{cleob} in both inclusive and exclusive studies of the reaction $\psi'\to\pi^0 h_c,\;h_c\to\gamma\eta_c$.  The inclusive $\pi^0$ recoil spectrum is illustrated in Fig. 4 (left), and the exclusive recoil spectrum obtained by summing six prominent decay channels of $\eta_c$ is shown in Fig. 4 (right).  The preliminary results are: \vspace{10pt}\\ 
\hspace*{0.5cm} Inclusive: $M(h_c)=3524.8\pm0.7$(stat)$\pm\sim1$(syst) MeV, significance $>3\sigma$.\\
\hspace*{0.5cm} Exclusive: $M(h_c)=3524.4\pm0.9$(stat) MeV, significance $>5\sigma$.
\vspace{10pt} \\
The preliminary result for product branching ratio, with a very liberal present error assignment, is $\mathcal{B}(\psi'\to\pi^0h_c)\times\mathcal{B}(h_c\to\gamma\eta_c)=(4\pm2)\times10^{-4}$.  Thus, the present CLEO result is that $\Delta M_{hf}(1P)=+0.66\pm0.55$(stat) MeV.  We note that the preliminary result presented at QWG III by E835 from their analysis of the reaction $p\bar{p}\to h_c \to \gamma\eta_c$ is $\Delta M_{hf}(1P)=-0.6\pm0.2$(stat)$\pm0.2$(syst) MeV \cite{e835a}.  We hope that in due time these different results can be reconciled.  In the meanwhile, it is clear that no large departures from the `naive' pQCD expectation that $\Delta M_{hf}=0$ are being observed.

\begin{figure}[t]
%\begin{center}
\setlength{\tabcolsep}{0pt}
\begin{tabular}{p{3.in}p{3.in}}
\includegraphics[width=2.8in]{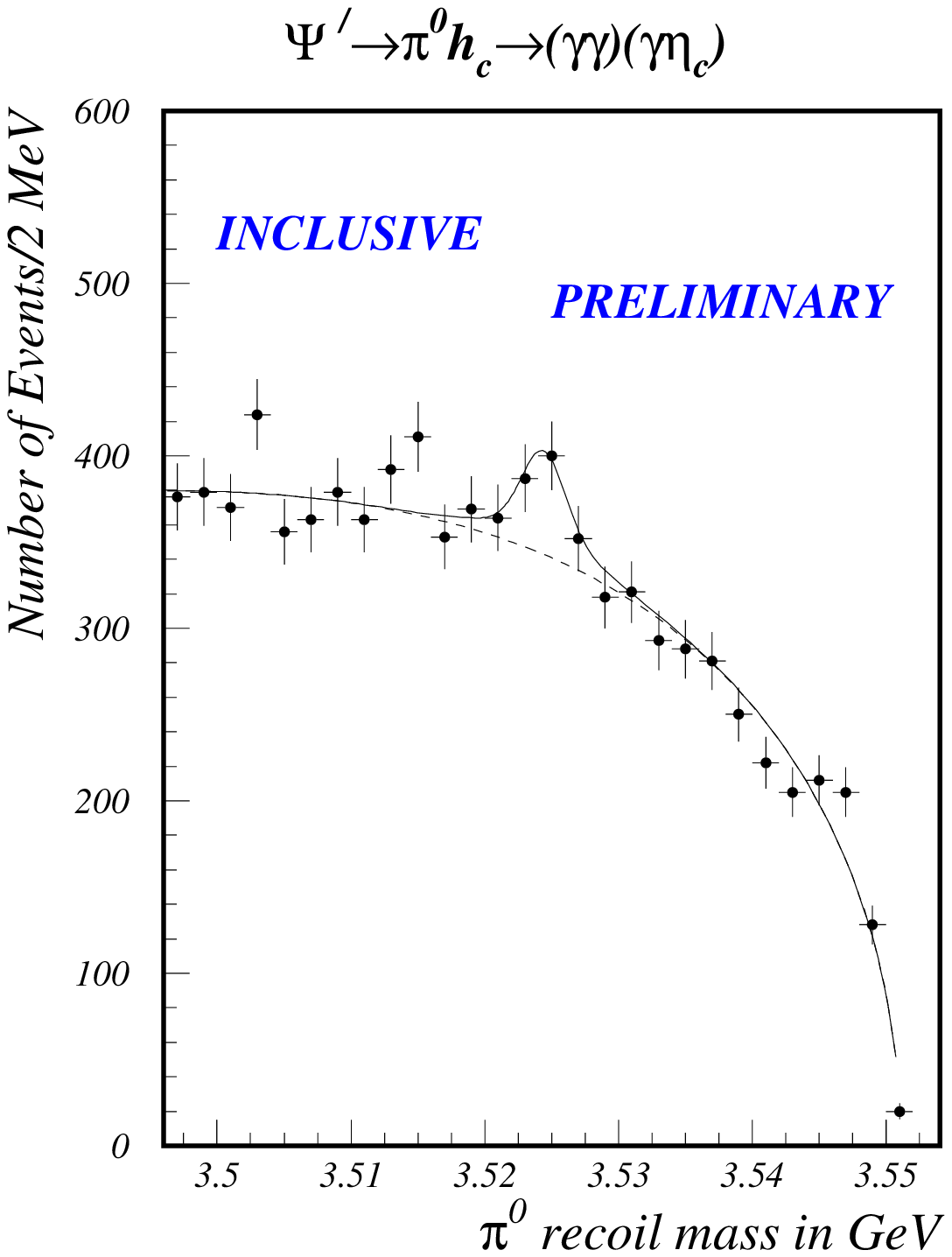}
& 
\vspace*{-8.cm}
\rotatebox{-90}{\includegraphics*[width=2.6in]{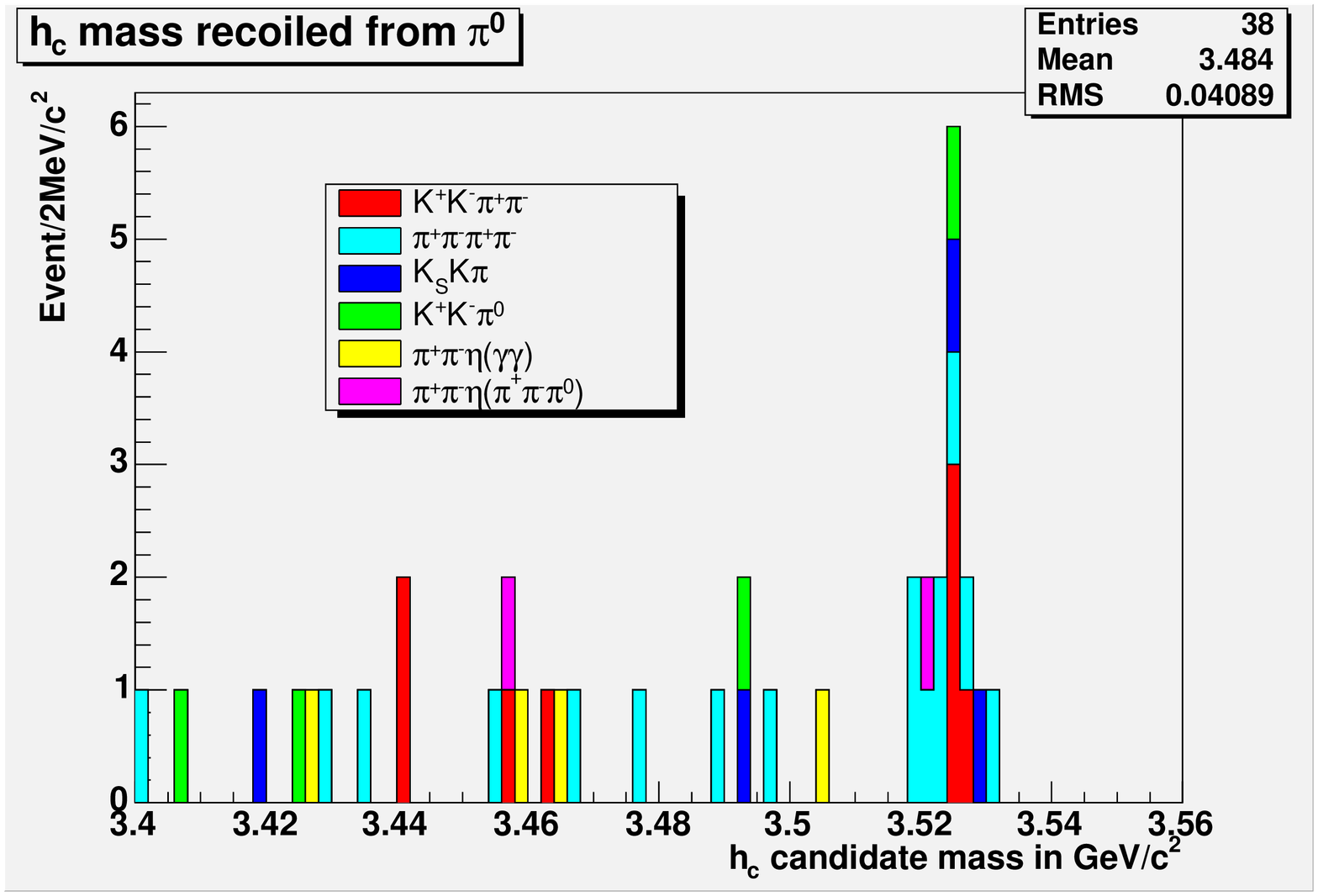}}
\\
\end{tabular}
%\end{center}
%\caption{CLEO results for $h_c$ from analysis of the inclusive reaction $\psi'\to\pi^0h_c\to(\gamma\gamma)(\gamma\eta_c)$.}
\caption{CLEO results for $h_c$ from analyses of the reaction $\psi'\to\pi^0h_c\to(\gamma\gamma)(\gamma\eta_c)$: (left) inclusive analysis; (right) exclusive analysis with $\eta_c$ decays into six hadronic channels.}
\end{figure}

%\begin{figure}[t]
%\begin{center}
%\includegraphics[width=3.in]{final_inc.eps} 
%\rotatebox{-90}{\includegraphics*[width=2.8in]{hcmassall.ps}}
%\end{center}
%\caption{CLEO results for $h_c$ from analysis of the inclusive reaction $\psi'\to\pi^0h_c\to(\gamma\gamma)(\gamma\eta_c)$.}
%\end{figure}

\subsection{An Uninvited Guest, X(3872)}

Last year the Belle Collaboration \cite{belleb} announced the arrival of an uninvited guest.  They reported the identification of a weak, but unambiguous signal for a narrow resonance in the reaction, $B^\pm\to K^\pm X(3872),\;X(3872)\to\pi^+\pi^-J/\psi$.  Its existence has since been confirmed by CDF, D\O, and BaBar, with the overall result, $M$(X)$=3872\pm1$ MeV, $\Gamma$(X)$<3$ MeV.  The mass, which does not easily fit in the charmonium spectrum, the narrow width of the resonance, and the fact that it has not been seen in any decay channel other than $\pi^+\pi^-J/\psi$, make it quite mysterious.  Since its $J^{PC}$ can not be determined from its sole observed decay, it provides very fertile ground for theoretical speculations about its nature.  Barnes et al \cite{barnes} and Eichten et al \cite{eichten} believe that with channel mixing and uncertainties in potential model calculations, it could still be a charmonium state, with $2^3P_{0,2}$ and $1^1D_2$ as favorites.  Because $M(D^0)+M(D^{0*})=3871.3\pm1.0$ MeV, Swanson \cite{swanson} and Tornqvist \cite{tornqvist} advocate that it is a weakly bound molecule, most likely with $J^{PC}=0^{-+}$.  Close and Page \cite{closepage} propose a $c\bar{c}$ hybrid possibility, and not to be left behind, I have suggested that it can be a vector glueball mixed with nearby $c\bar{c}$ vectors \cite{seth}.  In other words, almost anything is possible. The possibilities can only be limited by observing X(3872) in other decay channels, and determining, or at least limiting, its $J^{PC}$.  With this in mind, CLEO \cite{cleoc} has searched for X(3872) population in two-photon fusion, $\gamma\gamma\to\mathrm{X}(3872)\to\pi^+\pi^-J/\psi$, which would lead to positive charge conjugation for X(3872), and X(3872) population via initial state radiation (ISR) which would lead to $J^{PC}=1^{--}$.  As shown in Fig. 5, X(3872) is not observed in either reaction, and rather stringent 90\% confidence upper limits are established
$$(2J+1)\Gamma_{\gamma\gamma}(X(3872))\times\mathcal{B}(X\to\pi^+\pi^-J/\psi)<12.9\;\mathrm{eV}$$
$$\Gamma_{ee}(X(3872))\times\mathcal{B}(X\to\pi^+\pi^-J/\psi)<8.3\;\mathrm{eV}$$
We note that earlier Yuan et al \cite{besa} had analyzed BES data to obtain a similar limit for the ISR population of X(3872).

For the present, X(3872) remains a mystery!

%\begin{figure}[t]
%\begin{center}
%\rotatebox{-90}{\includegraphics*[width=2.8in]{hcmassall.ps}}
%\end{center}
%\caption{CLEO results for $h_c$ from analysis of the exclusive reaction $\psi'\to\pi^0h_c\to(\gamma\gamma)(\gamma\eta_c)$, with $\eta_c$ decays into six hadronic channels, as indicated.}
%\end{figure}

\begin{figure}[t]
\begin{center}
\includegraphics*[width=3.0in]{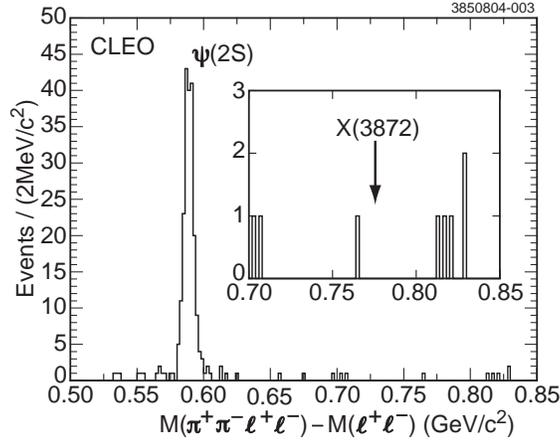}
\end{center}
\caption{Distribution of observed events as function of $\Delta M\equiv M(\pi^+\pi^-l^+l^-)-M(l^+l^-)$. The peak for $\psi(2S)$ from ISR is clearly visible, but no enhancement for X(3872) is observed.}
\end{figure}

\subsection{Radiative Decays of $\psi'(3686)$ and the $\chi_{cJ}$ States}

The radiative decays of $\psi'(3686)$ to $\chi_{cJ}$, and the subsequent radiative decay of $\chi_{cJ}$ to $J/\psi$ were primarily measured by the Crystal Ball [21,22], whose results used to dominate the earlier PDG compilations.  Recently, the PDG \cite{pdg} has started to make so called `global fits' of a variety of measurements of $\psi'$ and $\chi_{cJ}$ decays, with results which often differ considerably from the actual measurements.  At CLEO, new measurements have been made of the decays $\psi'\to\gamma\chi_J(J=0,1,2)$, the cascade decays $\psi'\to\gamma\chi_J\to\gamma\gamma J/\psi$, and the hadronic decays $\psi'\to\pi^0J/\psi$, and $\psi'\to\eta J/\psi$.  Some of these results are now available [24,25], and they show that `global fits' can be quite misleading (note, e.g., $\mathcal{B}(\psi'\to\gamma\chi_2)$ in Table 1).

\begin{table}[b]
\begin{center}
\begin{tabular}{l|l|l|l}
\hline
\% & PDG \cite{pdg} & CBALL \cite{cballc} & CLEO \cite{cleod} \\
\hline
$\mathcal{B}(\psi'\to\gamma\chi_0)$ & 8.6(7) & 9.9(9) & 9.22(47) \\
$\mathcal{B}(\psi'\to\gamma\chi_1)$ & 8.4(8) & 9.0(9) & 9.07(55) \\
$\mathcal{B}(\psi'\to\gamma\chi_2)$ & 6.4(6) & 8.0(9) & 9.33(63) \\
$\mathcal{B}(\psi'\to\gamma\eta_c)$ & 0.28(8) & 0.25(6) & 0.32(7) \\
\hline
\end{tabular}
\end{center}
\caption{Branching ratios (in \%) for $\psi'$ radiative transitions to $\chi_{cJ}$ and $\eta_c$.  The PDG results are from their global fit and differ markedly from the original Crystal Ball measurements.}
\end{table}

The original Crystal Ball measurements \cite{cballc} of cascades $\psi'\to\gamma\chi_J\to\gamma\gamma J/\psi$ have been recently supplemented by BES measurements \cite{besb}, and preliminary results from E835 \cite{e835b}.  

\begin{table}[b]
\begin{tabular}{l|l|l|l|l}
\hline
\% & PDG \cite{pdg} & E835 \cite{e835b} & BES \cite{besb} & CLEO \cite{cleoe} \\
\hline
$\mathcal{B}(\psi'\to e^+e^-)$ & 0.76(3) & 0.68(4) &  &  \\
$\mathcal{B}(\psi'\to J/\psi X)$ & 57.6(20) &  & 59.2(28) & 59.6(24)$^*$ \\
$\mathcal{B}(\psi'\to J/\psi \pi^0\pi^0)$ & 18.8(12) & 16.7(15) & 18.1(10) & 16.9(8)$^*$ \\
$\mathcal{B}(\psi'\to J/\psi \pi^+\pi^-)$ & 31.7(11) & 29.2(19) &  & 33.3(18)$^*$ \\
$\mathcal{B}(\psi'\to J/\psi \eta)$ & 3.16(22) & 2.8(3) & 2.98(25) & 3.3(1)$^*$ \\
$\mathcal{B}(\psi'\to J/\psi \pi^0)$ & 0.10(2) & 0.24(5) & 0.14(2) & 0.15(2)$^*$ \\
$\mathcal{B}(\psi'\to \gamma\chi_0\to\gamma\gamma J/\psi)$ & 0.10(1) & 0.13(5)$^*$ &  &  ? \\
$\mathcal{B}(\psi'\to \gamma\chi_1\to\gamma\gamma J/\psi)$ & 2.67(15) & 3.13(22)$^*$ & 2.81(24) &  ? \\
$\mathcal{B}(\psi'\to \gamma\chi_2\to\gamma\gamma J/\psi)$ & 1.30(8) & 1.91(16)$^*$ & 1.62(13) &  ? \\
$\mathcal{B}(\psi'\to \Sigma \gamma\gamma J/\psi)$ & \textbf{4.07(17)} & \textbf{5.17(28)$^*$} & \textbf{$\approx$4.5} & \textbf{6.2(11)$^*$} \\
$\mathcal{B}(\psi'\to \frac{\pi^0\pi^0 J/\psi}{\pi^+\pi^- J/\psi})$ & \textbf{0.59(4)} & \textbf{0.57(5)} & \textbf{0.57(3)} & \textbf{0.51(4)$^*$} \\
\hline
\end{tabular}
\caption{Latest results for hadronic and cascade decays of $\psi'$.}
\end{table}

The CLEO cascade data is being presently analyzed.  The final results are not yet available, but the quality of the cascade Dalitz plot shown in Fig. 6 \cite{cleog} assures us that higher precision results for all three $\chi_{cJ}$ resonances will be available soon. The available results for the cascade branching ratios, $\mathcal{B}_1(\psi'\to\gamma\chi_J\to\gamma\gamma J/\psi)$, and other hadronic decays of $\psi'$ are summarized in Table 2.

\begin{figure}
\begin{center}
\includegraphics*[width=3.5in]{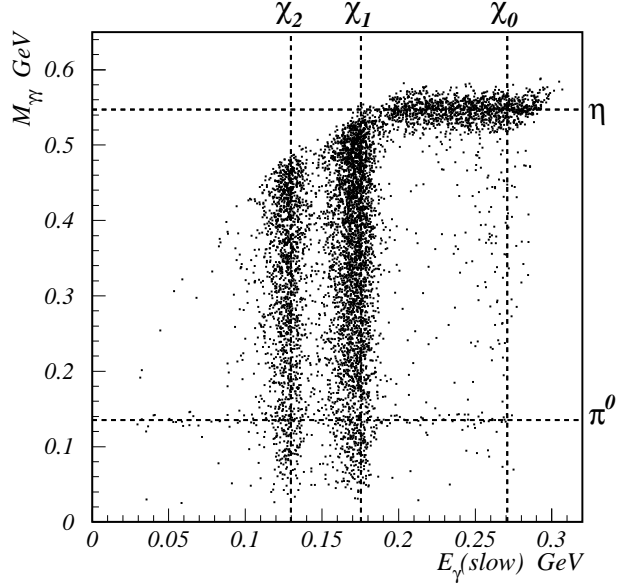}
\end{center}
\caption{Dalitz plot for the two photon decays $\psi'\to(\gamma\gamma)J/\psi$.  The bands for $\chi_{cJ}$ resonances and $\pi^0$ and $\eta$ are clearly visible \cite{cleog}.}
\end{figure}

\subsection{The Problem with $\mathcal{B}(\rho-\pi)$ (or the 13\% rule) of $\psi'$ and $J/\psi$}

According to pQCD, the partial widths for both the leptonic and hadronic decays of the vector resonances $J/\psi$ and $\psi'$ depend on their radial wave functions at the origin.  Therefore, one expects that the ratio
$$Q_{ee}\equiv\Gamma(\psi'\to e^+e^-)/\Gamma(J/\psi\to e^+e^-)=[\alpha^2_{em}|R(0)|^2_{\psi'}/[\alpha^2_{em}|R(0)|^2_{J/\psi},$$
which is experimentally known to be $0.13\pm0.01$, should be identical to $Q_{ggg}\equiv\Gamma(\psi'\to\sum LH)/\Gamma(J\psi\to\sum LH)$ (neglecting the variation of $\alpha_{strong}$ between $J/\psi$ and $\psi'$).  In fact, the ratio for the sum of all light hadron decays is $0.17\pm0.03$, as expected.  A naive extension of this identity is that the same ratio should be expected in individual hadronic decay channels.  However, already in 1983, Franklin et al. \cite{franklin} showed that the expectation was strongly violated in the decays of $J/\psi$ and $\psi'$ into vector/pseudoscalar pairs $\rho\pi$ and $K^*K^\pm$, the ratio being an order of magnitude smaller.  The $\rho-\pi$, or the V/PS, or the 13\% problem was born.

Over the years many theoretical explanations for the observed violation of the `13\% rule' have been offered involving glueballs, form factors, nodes, color octets, etc., but they all appear rather contrived, and there is no consensus.  On the experimental front, BES \cite{besc} has measured many other hadronic decays and published many papers since 1996, and shown that the ratio varies all over the map, and now just for V/PS pairs.  Recently, CLEO [25,31] has measured many two-body and many-body decays of $\psi'$. The combined results for the ratio are displayed in Fig. 7.  The only observation that one can make is that two isospin violating decays (to $\omega\pi^0$, $\rho\eta$), two isospin conserving decays ($b_1^0\pi^0$, $b_1^+\pi^+$), and several many-body decays appear to satisfy the rule, but most do not.  My personal feeling is that in expecting individual hadronic decays to follow this `rule', one is stretching pQCD too thin.

\begin{figure}
\begin{center}
\begin{tabular}{p{3.in}p{3.in}}
\includegraphics[width=2.8in]{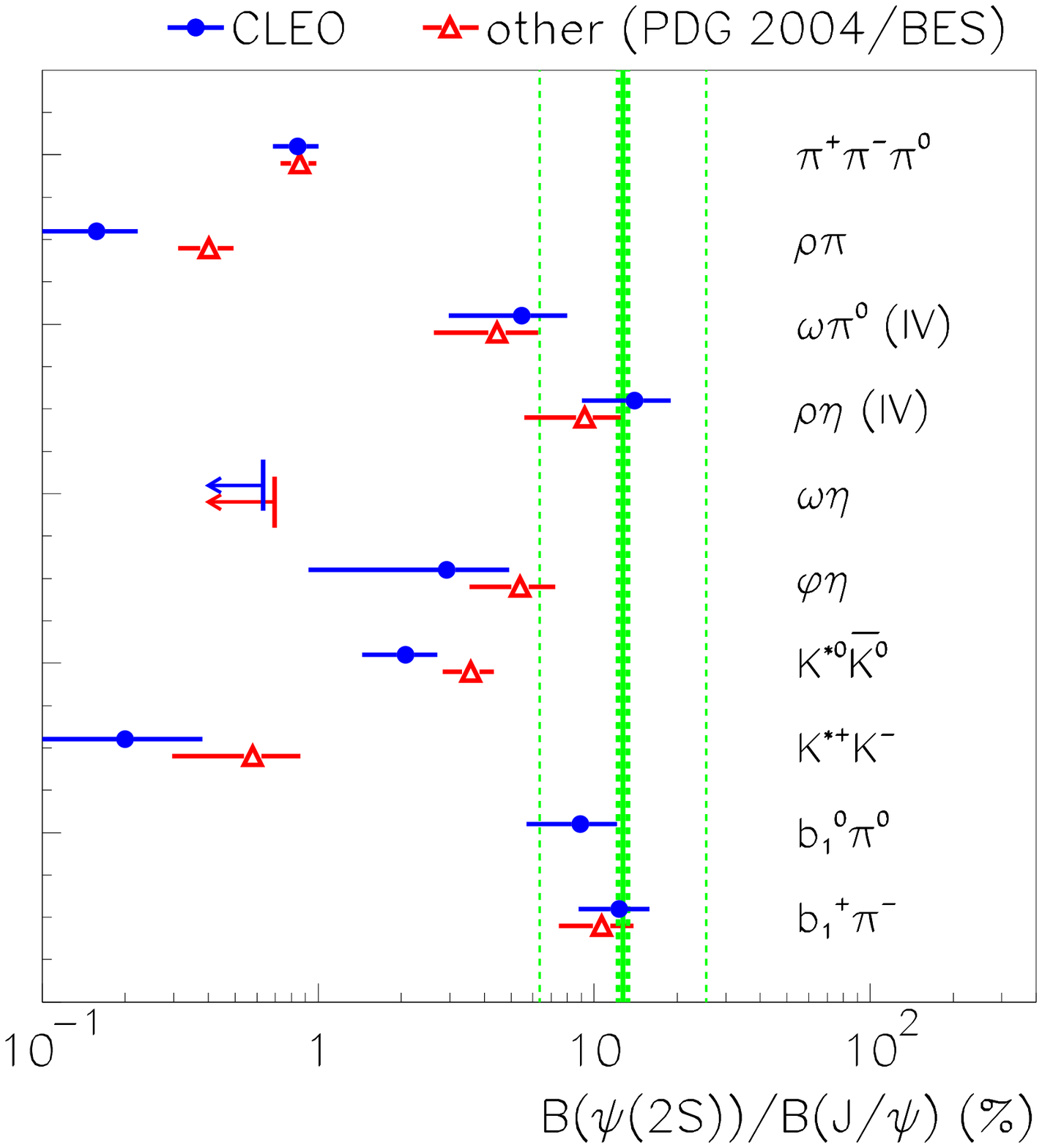}
&
\vspace*{-8.1cm}
\includegraphics[width=2.8in]{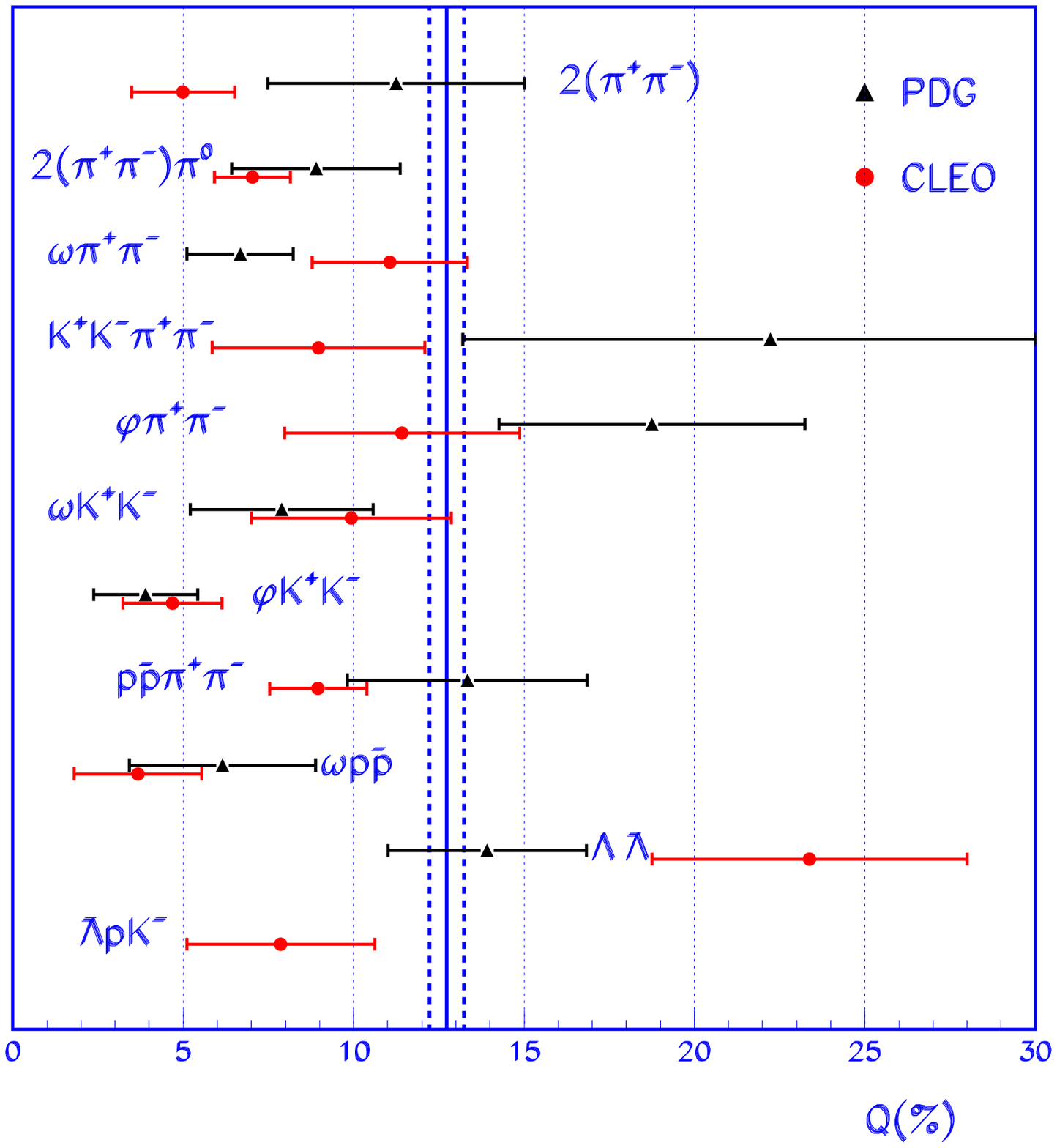}
\\
\end{tabular}
\vspace*{-1.8cm}
\end{center}
\caption{A compilation of the CLEO [25,31] and BES [30] results for the ratio $Q_{ggg}\equiv\mathcal{B}(\psi'\to LH)/\mathcal{B}(J/\psi\to LH)$.}
\end{figure}

\subsection{Charmonium Vectors above $D\bar{D}$ Threshold}

Above the $D\bar{D}$ threshold at 3740 MeV one expects to have the ($3^{3,1}S_J$), ($4^{3,1}S_J$), ($1^{3,1}D_J$)..., and ($2^{3,1}P_J$) states of charmonium.  Most of these can decay to $D\bar{D}$ and are expected to have large widths.  A few, such as $1^{3,1}D_2$, may be narrow.  However, only one state, the vector at 3770 MeV, called $\psi(3S)$, has been studied in some detail because it is a copious source of $D\bar{D}$.  For the rest, all we have had for a very long time are several measurements of the parameter $R\equiv\sigma(e^+e^-\to\mathrm{hadrons})/\sigma(e^+e^-\to\mathrm{muons})$, which have been notable  for their disagreement with each other.  One of them, by DASP \cite{dasp}, observed relatively narrow structures and reported parameters for three vector states, which have been adopted by PDG \cite{pdg} despite the fact that none of the other measurements observed the well-defined structures seen by DASP.  A recent detailed, high statistics measurement of $R$ by BES \cite{besd} has now helped resolve the situation.  It is found that the BES measurement is in excellent agreement with the Crystal Ball measurement \cite{cballe}.  The two datasets have been analyzed \cite{sethb} separately and found to yield nearly identical parameters for the three vector resonances.  The fits are shown in Fig. 8 and the resonance parameters are listed in Table 3.  It is found that the total and leptonic widths of these resonances are considerably larger than the DASP values listed in PDG \cite{pdg}.

\begin{figure}
\begin{center}
\includegraphics*[width=3.5in]{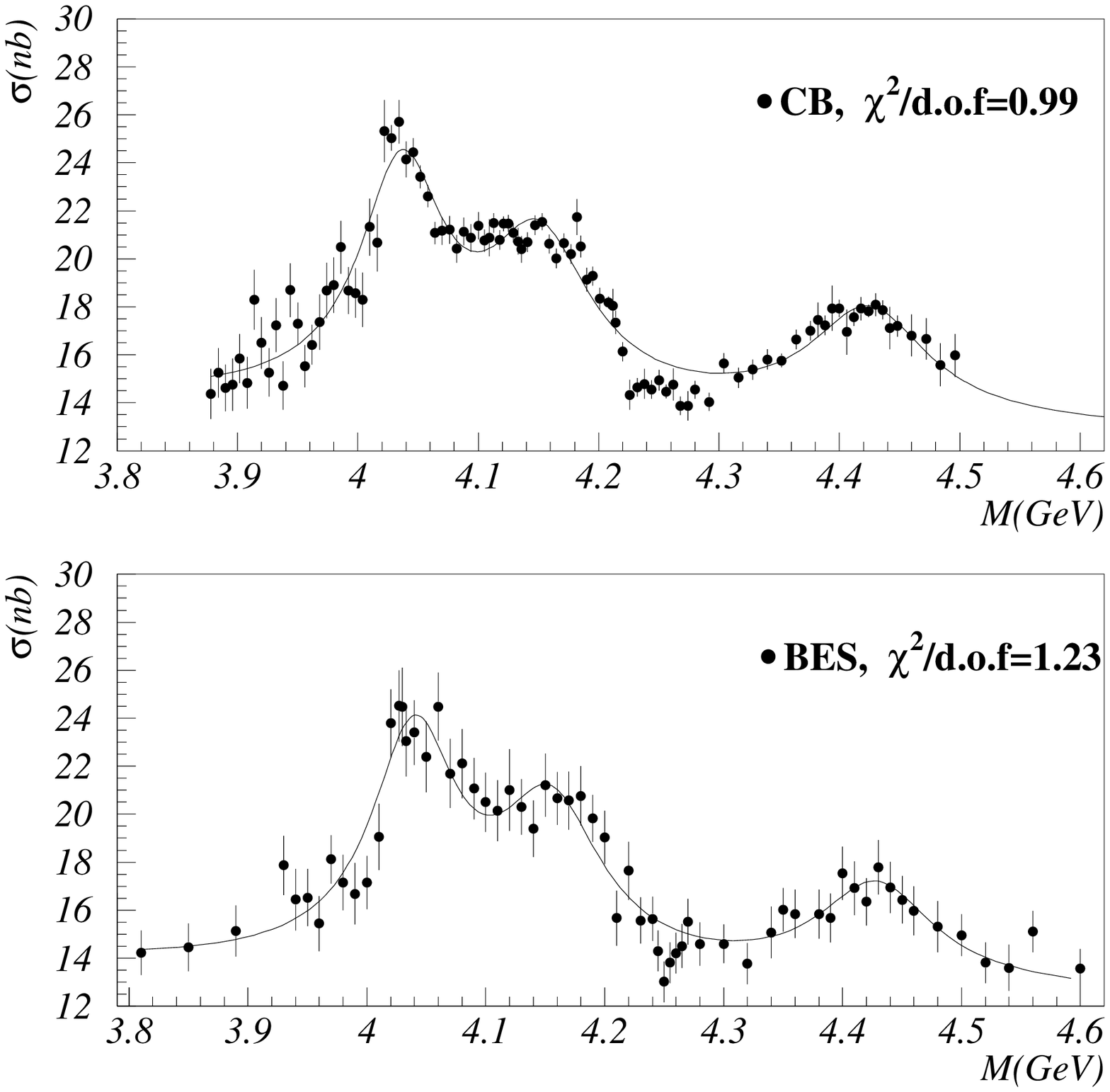}
\vspace*{-0.8cm}
\end{center}
\caption{$R\equiv\sigma(e^+e^-\to\mathrm{hadrons})/\sigma(e^+e^-\to\mathrm{leptons})$ from BES \cite{besd} and Crystal Ball \cite{cballe}.  The fits are from ref. \cite{sethb}.}
\end{figure}

\begin{table}[b]
\begin{center}
\begin{small}
\begin{tabular}{|c|c|c|c|}
\hline
\hline
 & $M^{(1)}$ & $\Gamma^{(1)}_{tot}$ & $\Gamma^{(1)}_{ee}$ \\
 & (MeV) & (MeV) & (keV) \\
\hline
PDG\cite{pdg} & $4040 \pm 10$ & $52 \pm 10$ & $0.75 \pm 0.15$ \\
CB\cite{cballe} & $4037 \pm 2$ & $85 \pm 10$ & $0.88 \pm 0.11$ \\
BES\cite{besd} & $4040 \pm 1$ & $89 \pm 6$ & $0.91 \pm 0.13$ \\
\hline
CB+BES & $4039.4 \pm 0.9$ & $88 \pm 5$ & $0.89 \pm 0.08$ \\
\hline
\hline
 & $M^{(2)}$ & $\Gamma^{(2)}_{tot}$ & $\Gamma^{(2)}_{ee}$ \\
\hline
PDG\cite{pdg} & $4159 \pm 20$ & $78 \pm 20$ & $0.77 \pm 0.23$ \\
CB\cite{cballe} & $4151 \pm 4$ & $107 \pm 10$ & $0.83 \pm 0.08$ \\
BES\cite{besd} & $4155 \pm 5$ & $107 \pm 16$ & $0.84 \pm 0.13$ \\
\hline
CB+BES & $4153 \pm 3$ & $107 \pm 8$ &  $0.83 \pm 0.07$\\
\hline
\hline
 & $M^{(3)}$ & $\Gamma^{(3)}_{tot}$ & $\Gamma^{(3)}_{ee}$ \\
\hline
PDG\cite{pdg} & $4415 \pm 6$ & $43 \pm 15$ & $0.47 \pm 0.10$ \\
CB\cite{cballe} & $4425 \pm 6$ & $119 \pm 16$ & $0.72 \pm 0.11$ \\
BES\cite{besd} & $4429 \pm 9$ & $118 \pm 35$ & $0.64 \pm 0.23$ \\
\hline
CB+BES & $4426 \pm 5$ & $119 \pm 15$ &  $0.71 \pm 0.10$\\
\hline
\hline
\end{tabular}
\end{small}
\end{center}
\caption{Resonance parameters of the three vector states of charmonium above the $D\bar{D}$ threshold. From \cite{sethb}.}
\end{table}

\subsection{Threshold Resonances in $J/\psi$ Decay}

With its huge sample of 58 million $J/\psi$, BES has been able to study several unique phenomena.  The most provocative among these are threshold baryon-antibaryon and meson-baryon resonances.

The first of these is the baryonium or the $p\bar{p}$ resonance.  This was the hot topic of the early low-energy antiproton experiments at Brookhaven and LEAR (CERN).  No convincing evidence was found and the fever subsided.  Now, BES \cite{bese} has provided evidence for a threshold enhancement in the reaction $J/\psi\to\gamma+(p\bar{p})$, which is illustrated in Fig. 9.  They claim that it can not be explained in terms of phase space, and propose that it reflects the tail of a resonant state of $p\bar{p}$ baryonium (bound by about 25 MeV) with mass $M(p\bar{p})=1859^{+3}_{-10}$$^{+5}_{-25}$ MeV, and width $\Gamma<30$ MeV.  For such an exciting claim independent confirmation is definitely needed.

\begin{figure}
\begin{center}
\scalebox{0.8}{\includegraphics{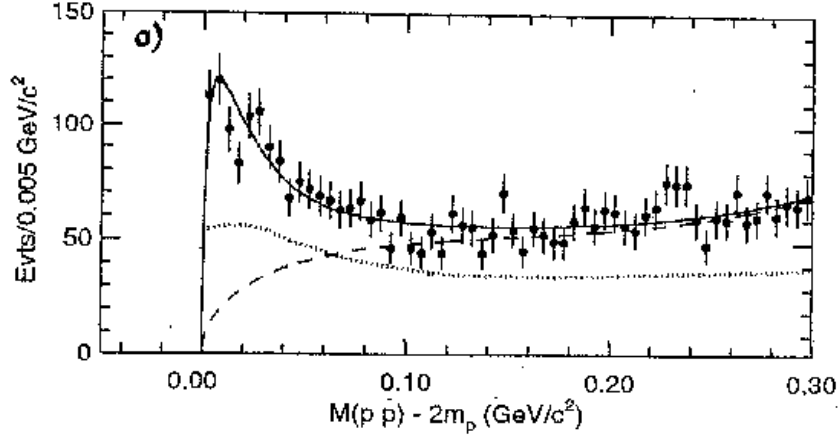}}
\end{center}
\caption{Invariant mass difference $M(p\bar{p})-2m_p$ for the reaction $J/\psi\to\gamma+(p\bar{p})$. The solid curve is the fit for a bound $p\bar{p}$ resonance [36].}
\end{figure}

More recently, BES \cite{besf} has claimed observation of a small enhancement in the $p\bar{\Lambda}$ invariant mass in the reaction $J/\psi\to K^-+(p\bar{\Lambda})$ which they interpret as a slightly unbound (by $\sim20$ MeV) $p\bar{\Lambda}$ resonance with mass $M(p\bar{\Lambda})=2075\pm13$ MeV, and width $\Gamma=90\pm36$ MeV.  Going one step further, at QWG III they have also reported \cite{besg} an enhancement in the invariant mass of $K^-\bar{\Lambda}$ in the same reaction, and proposed that it represents an unbound state of $K^-\bar{\Lambda}$ with mass $M(K^-\bar{\Lambda})=1500-1600$ MeV, and width $\Gamma=70-110$ MeV.

If these interpretations are correct, and I must confess to personal reservations, we have here the beginning of a new, and rather unexpected, spectroscopy.

\subsection{Light-Quark Scalars in $J/\psi$ Decay}

BES [38,39] has studied the population of the well known light quark scalars $f_0(980)$, $f_0(1350)$, $f_0(1500)$, $f_0(1710)$.  Of particular note is their observation \cite{besh} of $f_0(600)$ or $\sigma$ in the reaction $J/\psi\to\omega+(\pi^+\pi^-)_\sigma$.  They report $M(\sigma)=541\pm39$ MeV, $\Gamma=504\pm84$ MeV.  The evidence for its excitation is shown by the shaded area in Fig. 10 (the large narrow peak is due to $f_2(1270)$).  At QWG III \cite{besg} BES also presented evidence for the so-called kappa $(\kappa)$ resonance in the reaction $J/\psi\to \bar{K}^* + (K^+\pi^-)_\kappa$.  They find $M(\kappa)=841\pm78^{+81}_{-73}$ MeV, and $\Gamma(\kappa)=618\pm182^{+96}_{-144}$ MeV.

As mentioned earlier, these discoveries are important, and need to be confirmed.  We hope that CLEO-c can confirm them when it acquires its expected larger datasets at $J/\psi$ and $\psi'$.

\newpage

\begin{figure}
\begin{center}
\scalebox{0.8}{\includegraphics{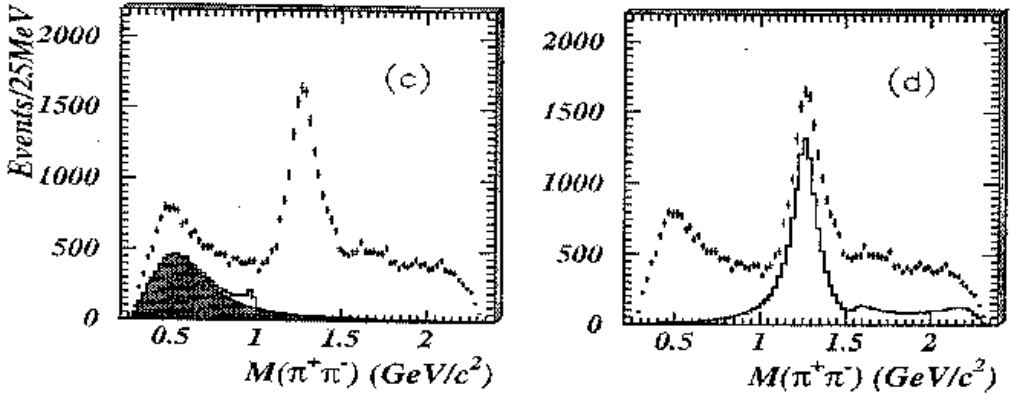}}
\end{center}
\caption{Plot of two pion invariant mass in the reaction $J/\psi\to\omega+(\pi^+\pi^-)$.  The shaded area represents the $\sigma$ resonance.  The large peak is due to $f_2(1270)$ [39].}
\end{figure}

\section{New Results in Bottomonium}

As shown in Fig. 2, CLEO III has acquired large amounts of data at the bound Upsilon resonances $\Upsilon(1S)$, $\Upsilon(2S)$ and $\Upsilon(3S)$.  This has enabled it to make many new high precision measurements in the spectroscopy of $b\bar{b}$ bottomonium.

\subsection{Leptonic Decays of $\Upsilon(1S,2S,3S)$}

Previous measurements of $\mathcal{B}(\Upsilon(nS)\to\mu^+\mu^-)$ were made by scanning the resonances and making simultaneous fits to the leptonic and hadronic excitation functions.  CLEO \cite{cleoh} has now made precision determinations of these branching ratios by measuring the ratios of the leptonic and hadronic yields at (or near) the peaks only.  The new results for $\mathcal{B}_{\mu\mu}$ are shown in Table 4.  It is to be noted that while the new result for $\mathcal{B}_{\mu\mu}(\Upsilon(1S))$ is in good agreement with the earlier measurements as summarized by PDG, the new CLEO results for $\mathcal{B}_{\mu\mu}(\Upsilon(2S))$ and $\mathcal{B}_{\mu\mu}(\Upsilon(3S))$ are $\sim55\%$ and $32\%$ larger, respectively.  When combined with the existing measurements of $\Gamma_{\mu\mu}\Gamma_h/\Gamma$ the new measurements lead to similar increases in the total widths of $\Upsilon(2S)$ and $\Upsilon(3S)$.

\begin{table}[b]
\begin{center}
\begin{tabular}{rcc}
\hline
 & $B_{\mu\mu}(\%)$ CLEO & $B_{\mu\mu}(\%)$ PDG\\
\hline
$\Upsilon(1S)$ & $2.49\pm0.02\pm0.07$ & $2.48\pm0.06$ \\
$\Upsilon(2S)$ & $2.03\pm0.03\pm0.08$ & $1.31\pm0.21$ \\
$\Upsilon(3S)$ & $2.39\pm0.07\pm0.10$ & $1.81\pm0.17$ \\
\hline
\end{tabular}
\end{center}
\caption{New results from CLEO for the leptonic branching ratios [40].}
\end{table}

\subsection{Radiative Decays of $\Upsilon(2S)$ and $\Upsilon(3S)$}

CLEO \cite{cleoi} has made new measurements of the radiative transitions $\Upsilon(2S)\to\gamma\chi_{bJ}(1P)$ and $\Upsilon(3S)\to\gamma\chi_{bJ}(2P)$.  The masses of both sets of $\chi_{bJ}(J=0,1,2)$ states are found to be in excellent agreement with earlier results, as summarized by PDG \cite{pdg}.  The branching ratios $\mathcal{B}(\Upsilon(2S)\to\gamma\chi_{bJ}(1P))$ are similarly in excellent agreement with those in the literature, but $\mathcal{B}(\Upsilon(3S)\to\gamma\chi_{bJ}(2P))$ are all found to be smaller than those in PDG \cite{pdg}, by factors $0.79\pm0.19(\chi_{b0})$, $0.78\pm0.06(\chi_{b1})$, $0.72\pm0.06(\chi_{b2})$.

\subsection{First Observation of Bottomonium $^3D_2$ State}

Until now the only bottomonium states which had been firmly identified were the $n^3S_1$ and $n^3P_J$ $(n=1,2,3)$.  In a tour-de-force measurement of a four photon cascade, CLEO \cite{cleoj} has identified the $1^3D_2$ state of bottomonium, with $M(1^3D_2)=10161.1\pm0.6\pm1.6$ MeV.  The cascade was, $\Upsilon(3S)\to\gamma_1\chi(2^3P_J)\to\gamma_1\gamma_2\Upsilon(1^3D_2)\to\gamma_1\gamma_2\gamma_3\chi(1^3P_J)\to\gamma_1\gamma_2\gamma_3\gamma_4\Upsilon(1^3S_1)$.

\subsection{First Observation of a $\chi_b$ Hadronic Transition}

Until now the only hadronic transitions ever observed between bottomonium states were the two pion transitions $\Upsilon(nS)\to\Upsilon(n'S)+\pi\pi$.  Recently, a first was achieved by the CLEO \cite{cleok} identification of weak transitions $\chi_b(2P)\to\omega\Upsilon(1S)$, with $\mathcal{B}(\chi_{b1}(2P)\to\omega\Upsilon(1S))=(1.68\pm0.38)\%$, and $\mathcal{B}(\chi_{b2}(2P)\to\omega\Upsilon(1S))=(1.10\pm0.34)\%$.

\section{Postscript}

With CESR-III and CLEO-III transmuting into CESR-c and CLEO-c, we do not expect any new data to be taken for bottomonium spectroscopy, and there is little prospect of Belle or BaBar running at any but $\Upsilon(4S)$. However, we can expect CLEO to mine all it can from the existing body of their bottomonium data.  The prospects for spectroscopy in the charm/charmonium region are much brighter.  CLEO-c is dedicated to producing vast quantities of $D$ physics and charmonium physics during the next three years or so.  The BEPC-II (BES III) and GSI (PANDA) should then come online, and we should see many years of dedicated spectroscopy of the charm/charmonium region.  Keep tuned.

This research was supported in part by the US Department of Energy.


\section*{References}

\begin{thebibliography}{99}

\bibitem{cball}   CBALL, C. Edwards et al., \textit{Phys. Rev. Lett.} \textbf{48}(1982)70.

\bibitem{e760e835}  E760 Collaboration, T. A. Armstrong, et al., \textit{Phys. Rev.} \textbf{D 52}(1995)4839; E835 Collboration, M. Ambrogniani, \textit{Phys. Rev.} \textbf{D 64}(2001)052003.

\bibitem{bellea}  Belle Collaboration, S. K. Choi et al.,\textit{Phys. Rev. Lett.} \textbf{89}(2002)102001.

\bibitem{bellea'} Belle Collaboration, S. K. Choi et al., \textit{Phys. Rev. Lett.} \textbf{89}(2002)142001.

\bibitem{cleoa} CLEO Collaboration, D. M. Asner et al., \textit{Phys. Rev. Lett.} \textbf{92}(2004)142001.

\bibitem{babara} BaBar Collaboration, B. Aubert et al., \textit{Phys. Rev. Lett.} \textbf{92}(2004)142002.

\bibitem{cballb} E. D. Bloom and C. W. Peck, \textit{Ann. Rev. Nucl. Part. Sci.} \textbf{33}(1983)143.

\bibitem{e760a} E760 Collaboration, T. A. Armstrong et al., \textit{Phys. Rev. Lett.} \textbf{69}(1992)2337.

\bibitem{davethesis} D. Joffe, Ph. D. dissertation, Northwestern University, 2004, unpublished.

\bibitem{cleob} A. Tomaradze, QWG III Workshop, Beijing, Oct. 2004; also A. Tomaradze, elsewhere in these proceedings.

\bibitem{e835a} C. Patrignani, QWG III Workshop, Beijing, Oct. 2004.

\bibitem{belleb} Belle Collaboration, S. K. Choi et al., \textit{Phys. Rev. Lett.} \textbf{91}(2003)262001.

\bibitem{barnes} T. Barnes and S. Godfrey, \textit{Phys. Rev.} \textbf{D 69}(2004)054008.

\bibitem{eichten} E. J. Eichten, K. Lane, and C. Quigg, \textit{Phys. Rev. Lett.} \textbf{89}(2002)162002; \textit{Phys. Rev.} \textbf{D 69}(2004)094019.

\bibitem{swanson} E. S. Swanson, \textit{Phys. Lett.} \textbf{B 588}(2004)189; hep-ph/0406080.

\bibitem{tornqvist} N. A. T\"{o}rnqvist, \textit{Phys. Lett.} \textbf{B 590}(2004)209.

\bibitem{closepage} F. E. Close and P. R. Page, \textit{Phys. Lett.} \textbf{B 578}(2004)119.

\bibitem{seth} K. K. Seth, hep-ph/0411122.

\bibitem{cleoc} CLEO Collaboration, S. Dobbs et al., \textit{Phys. Rev. Lett.} (accepted); hep-ex/0410038; also P. Zweber, elsewhere in these proceedings.

\bibitem{besa} C. Z. Yuan, X. H. Mo, P. Wang, \textit{Phys. Lett.} \textbf{B 579}(2004)74.

\bibitem{cballc} M. J. Oreglia et al., \textit{Phys. Rev.} \textbf{D 25}(1982)2259.

\bibitem{cballd} J. Gaiser et al., \textit{Phys. Rev.} \textbf{D 34}(1986)711.

\bibitem{pdg} PDG04 Collaboration, S. Eidelman et al., \textit{Phys. Lett.} \textbf{B 592}(2004)1.

\bibitem{cleod} CLEO Collaboration, S. B. Athar et al., \textit{Phys. Rev.} \textbf{D} (accepted).

\bibitem{cleoe} B. Heltsley, QWG III, Beijing, Oct. 2004.

\bibitem{besb} BES Collaboration, J. Z. Bai, \textit{Phys. Rev.} \textbf{D 70}(2004)012006; M. Ablikin, \textit{Phys. Rev.} \textbf{D 70}(2004)012003.

\bibitem{e835b} E835 Collaboration, M. Andreotti et al., \textit{Phys. Rev.} \textbf{D}, submitted (2004); C. Patrignani, \textit{Proc. Hadron '03}, AIP Conf. Proc. 717 (2004) 581.

\bibitem{cleog} K. K. Seth, Proc. LATTICE '04 (Fermilab), June 2004, in press.

\bibitem{franklin} M. E. B. Franklin et al., \textit{Phys. Rev. Lett.} \textbf{51}(1983)963.

\bibitem{besc} BES Collaboration, J. Z. Bai et al., \textit{Phys. Rev.} \textbf{D 54}(1996)1221; \textit{Phys. Rev. Lett.} \textbf{81}(1998)5080; \textit{Phys. Rev. Lett.} \textbf{83}(1999)1918; \textit{Phys. Rev.} \textbf{D 67}(2003)052002.

\bibitem{cleof} CLEO Collaboration, N. E. Adam et al., \textit{Phys. Rev. Lett.} (accepted); also H. Mahlke-Krueger, elsewhere in these proceedings.

\bibitem{dasp} DASP Collaboration, R. Brandelik et al., \textit{Z. Phys.} \textbf{C1}(1979)233.

\bibitem{besd} BES Collaboration, J. Z. Bai et al., \textit{Phys. Rev. Lett.} \textbf{88}(2002)101802.

\bibitem{cballe} CBALL, A. Osterfeld et al., SLAC-PUB-4160(1986).

\bibitem{sethb} K. K. Seth, \texttt{hep-ex/0405007}.

\bibitem{bese} BES Collaboration, J. Z. Bai et al., \textit{Phys. Rev. Lett.} \textbf{91}(2003)022001.

\bibitem{besf} BES Collabortation, M. Ablikim et al., \textit{Phys. Rev. Lett.} \textbf{93}(2004)112002.

\bibitem{besg} X. Shen, QWG III, Beijing, Oct. 2004.

\bibitem{besh} BES Collaboration, M. Ablikim et al., \textit{Phys. Lett.} \textbf{B 598}(2004)149.

\bibitem{cleoh} CLEO Collaboration, G. S. Adams et al., \textit{Phys. Rev. Lett.} (accepted); hep-ex/0409027; also I. Danko, elsewhere in these proceedings.

\bibitem{cleoi} CLEO Collaboration, M. Artuso et al., \textit{Phys. Rev. Lett.} (accepted); hep-ex/0411068.

\bibitem{cleoj} CLEO Collaboration, G. Bonvicini et al., \textit{Phys. Rev.} \textbf{D 70}(2004)032001.

\bibitem{cleok} CLEO Collaboration, D. Cronin-Hennessy et al., \textit{Phys. Rev. Lett.} \textbf{92}(2004)222002.

\end{thebibliography}
\end{document}